\documentclass[oldversion]{aa}

\usepackage{times}
\usepackage{graphicx}
\usepackage{xspace}
\usepackage{epsfig}
\usepackage{natbib}
\usepackage{rotating}
\usepackage{dcolumn}

\begin{document}

\title{The X-ray emission from Z\,CMa during an FUor-like outburst \newline and the detection of its X-ray jet}

\author{B. Stelzer \inst{1} \and S. Hubrig \inst{2} \and S. Orlando \inst{1} \and G. Micela \inst{1} \and Z. Mikul\'a\v{s}ek \inst{3,4} \and M. Sch\"oller \inst{5}}

\offprints{B. Stelzer}

\institute{INAF - Osservatorio Astronomico di Palermo,
  Piazza del Parlamento 1,
  I-90134 Palermo, Italy \\ \email{B. Stelzer, stelzer@astropa.unipa.it} \and
  European Southern Observatory,
  Casilla\,19001,
  Santiago\,19, Chile \and 
  Department of Theoretical Physics and Astrophysics, Masaryk University, Kotlarska\,2, 61137 Brno, Czech Republic \and 
  Observatory and Planetarium of Johann Palisa, V\v{S}B--TU Ostrava, Czech Republic \and
  European Southern Observatory,
  Karl-Schwarzschild-Str. 2, 85748 Garching, Germany } 


\date{Received $<$29-01-2009$>$ / Accepted $<$16-03-2009$>$}

\abstract{
Accretion shocks have been recognized as important X-ray emission mechanism 
for pre-main sequence stars. Yet the X-ray properties of FUor outbursts, events that are caused by
violent accretion, have been given little attention. We have observed the FUor
object Z\,CMa during optical outburst and quiescence with {\em Chandra}. No significant
changes in X-ray brightness and spectral shape are found, suggesting that the X-ray 
emission is of coronal nature. Due to the binary nature of Z\,CMa
the origin of the X-ray source is ambiguous. However, the moderate hydrogen column
density derived from our data makes it unlikely that the embedded primary star
is the X-ray source. The secondary star, which is the FUor object,  
is thus responsible for both the X-ray emission
and the presently ongoing accretion outburst, which seem however to be unrelated phenomena. 
The secondary is also known to drive a large outflow and jet, that we detect here for the 
first time in X-rays. The distance of the X-ray emitting outflow 
source to the central star is higher than in jets of low-mass stars. 
}
{
}
{
}
{
}
{
}
{
}

\keywords{X-rays: stars -- Accretion -- stars: pre-main sequence -- stars: variables: general -- stars: activity -- stars: winds, outflows}

\maketitle

\section{Introduction}\label{sect:intro}

Variability is a dominant observational signature in pre-main sequence (pre-MS) stellar evolution. 
Besides the ubiquitous short-duration flares related to magnetic reconnection events,
various types of long-term outbursts are reported: 
FUor events are characterized by optical intensity changes of up to $4$\,mag and a fading
phase of decades, 
EXor events are less extreme and shorter (months to few years). 
Both types of outburst are associated with a sudden increase of the accretion rate,
such that the disk outshines the central star leading to characteristic spectral signatures
\citep[e.g.][]{Hartmann96.1}. 
The spectra of FUor objects resemble that of F-G supergiants, while EXor objects 
are of later spectral type. 
The FUor and EXor phenomena are probably recurrent,  
about once every $10^4$\,yrs for the FUors and every few years in the case of EXors. 
Only a minor fraction of pre-MS stars is classified as FUor or EXor \citep{Abraham04.1, Herbig08.1}.
Different mechanisms have been proposed as triggers for the outbursts:
dynamical interaction with a close binary companion 
\citep{Bonnell92.1, Reipurth04.1}, thermal instability in a disk with high accretion
from a surrounding envelope \citep{Bell94.1}, and 
changes in the magnetic field configuration \citep{vandenAncker04.1}. 

In recent years accretion has been recognized to make an important contribution to the X-ray emission 
of pre-MS stars \citep{Kastner02.1, Stelzer04.3}, making FUor and EXor objects interesting targets 
for X-ray studies. Plasma temperatures of up to a few $10^6$\,K can be produced in the accretion shocks 
that form when matter is funneled along the magnetic field lines onto the stellar surface.  
In the first dedicated X-ray survey of FUor objects 
\cite{Skinner09.1} have detected two of four stars with {\em XMM-Newton}. 
None of their targets were in a state of recent optical outburst during the X-ray observation. 
Only two EXors have been observed in X-rays during an optical outburst, with contradictory results
\citep{Kastner06.1, Audard05.2}.

The X-ray spectrum of the prototype FU\,Ori 
shows surprisingly a complex absorption pattern: 
While the harder emission is strongly absorbed and can be ascribed to an
embedded stellar corona, the origin of the weakly absorbed soft emission is unclear \citep{Skinner06.1}. 
Possible scenarios include
the overlaid effect of a binary companion, accretion shocks, and shocked jets. 
Indeed, \cite{Guenther09.1} showed recently that the unresolved soft component in the X-ray spectrum of the 
pre-MS star DG\,Tau 
can be explained as emission from a post-shock cooling zone in the innermost part of its optical outflow. 
Soft emission (few MK)
is only an indirect means of inferring outflows in X-rays and a signature that is easily confused with 
contributions from accretion.  
Direct detection of X-ray emission from pre-MS jets 
by means of a displacement with respect to the central coronal source 
has been achieved only in a handful of cases \citep[see summary in][]{Bonito07.1}. 

In this article we examine the X-ray properties of Z\,CMa during its recent optical outburst, 
we compare them to its quiescent state, and we present the X-ray detection of its jet. 

Z\,CMa is a $0.1^{\prime\prime}$ pre-MS binary. 
While the south-east (SE) FUor object dominates the light at optical wavelengths, 
the north-west (NW) component is a powerful infrared (IR) source \citep{Koresko91.1}. 
The FUor star has $\sim 3\,{\rm M_\odot}$ and is located just below the birthline in the
HR diagram \citep{Hartmann89.3}. Assuming that the IR source is 
coeval, it can be modeled as a 
B0\,III star with $16\,{\rm M_\odot}$ \citep{vandenAncker04.1}. 
The optically dominating component is, therefore, the secondary in the binary system.
This star is likely responsible both for the FUor phenomena and for the jet and molecular outflow 
observed at radio and optical wavelengths \citep{Poetzel89.1, Evans94.1, Velazquez01.1}. 
The long-term lightcurve of Z\,CMa exhibits features of both FUor 
and EXor-like events: Optical outbursts of $\sim 1$\,mag are superposed on 
a two decade long decay \citep{vandenAncker04.1}. 
An alternative explanation for the irregular light variations of Z\,CMa is scattered light
from the embedded primary that penetrates an envelope of variable thickness \citep{Hartmann96.1}.

\begin{figure}
\begin{center}
\includegraphics[width=7cm]{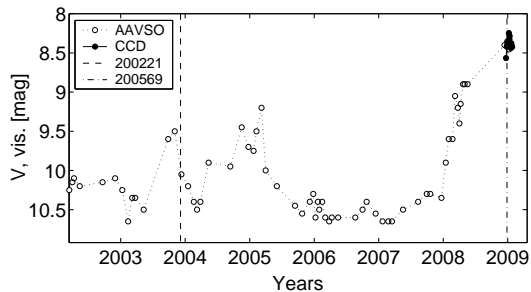}
\caption{Lightcurve of Z\,CMa; 
vertical lines indicate the date of {\em Chandra} observations.}
\label{fig:opt_lc}
\end{center}
\end{figure}

In Jan\,2008 Z\,CMa started a large optical outburst (see Fig.~\ref{fig:opt_lc}). 
Visual brightness estimates\footnote{We obtained the visual photometry from the American Association of
Variable Star Observers (AAVSO) at http://www.aavso.org} are available for the initial 
gradual increase by $\sim 1.5$\,mag.
Further visual observations and additional 
$BVR_{C}I_{C}$ CCD photometric observations\footnote{The CCD photometry was obtained by Czech observers L. Br\'at and L. \v{S}melcer; see http://var2.astro.cz/meduza/light-curves-ccd.php?star=Z\%20CMa\&shv=CMa of VSES CAS.} 
have resumed after a $\sim 6$ month-long gap. Z\,CMa reached 
its (temporary) maximum of $\sim 8.3$\,mag at the beginning of Jan 2009.  

\begin{figure}[t]
\begin{center}
\parbox{6.0cm}{
\includegraphics[width=6cm]{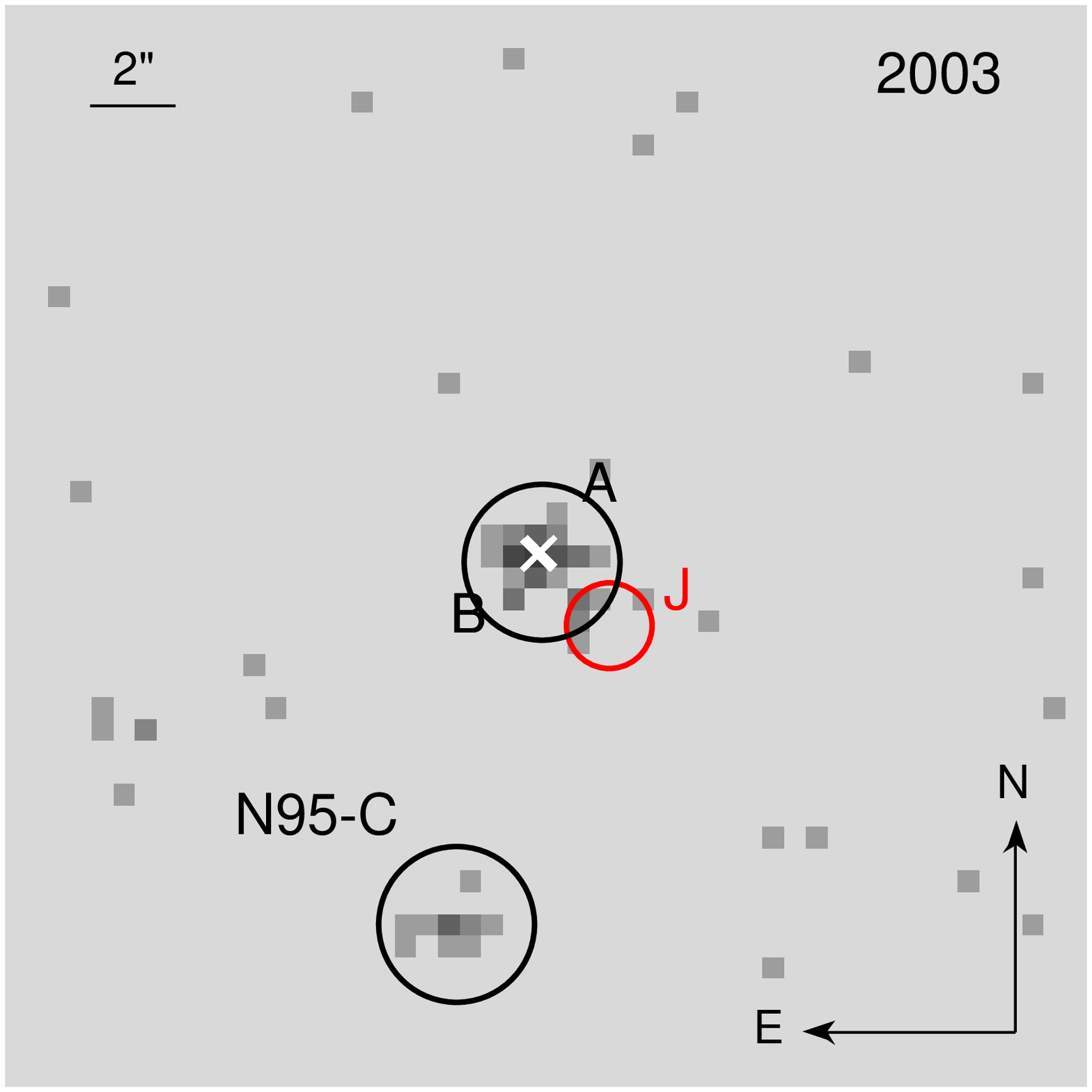}
}
\parbox{2.9cm}{
\vspace*{-1cm}
\hspace*{-1.5cm} \includegraphics[width=3.5cm,clip]{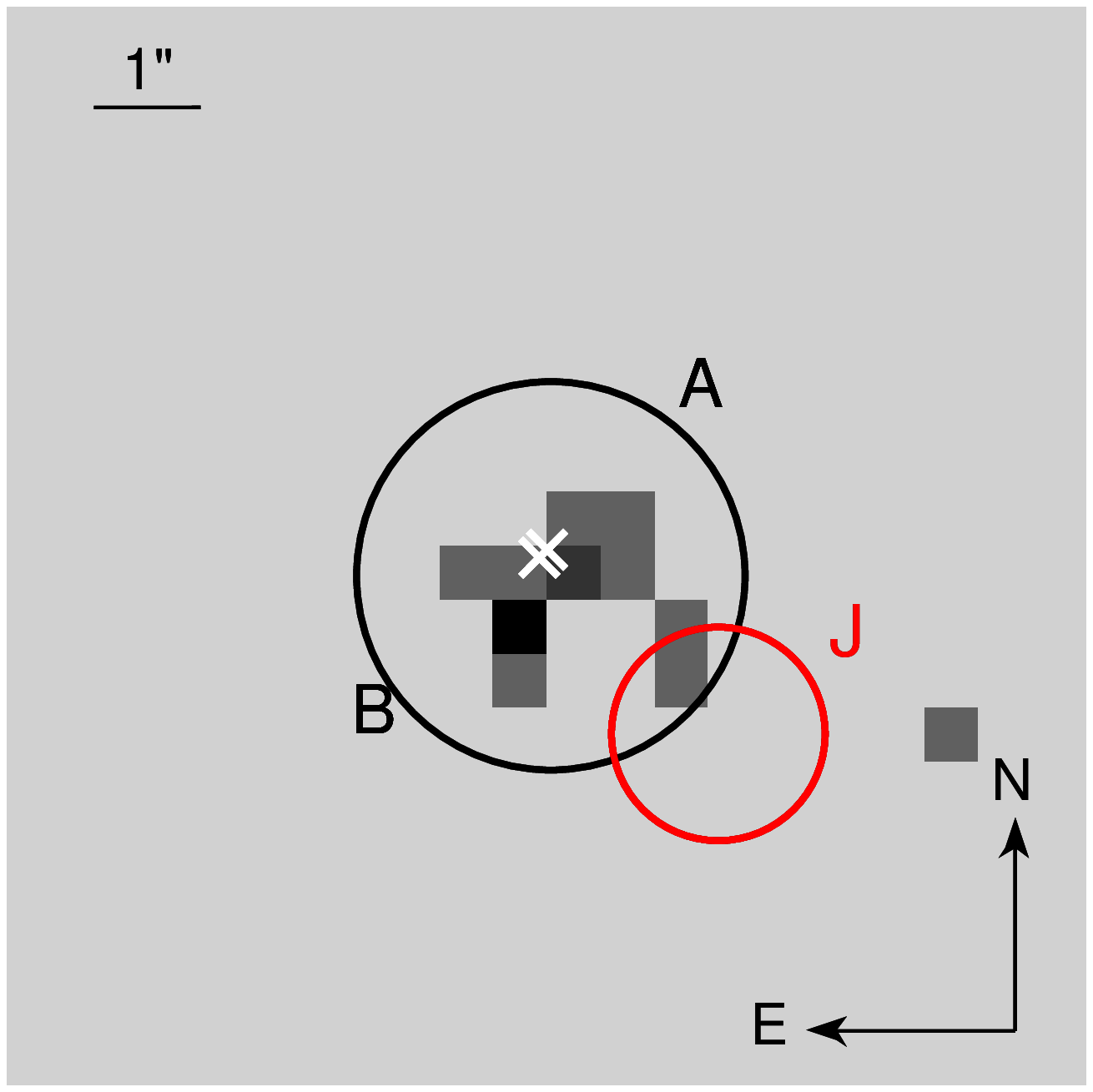}
}
\parbox{6.0cm}{
\includegraphics[width=6cm]{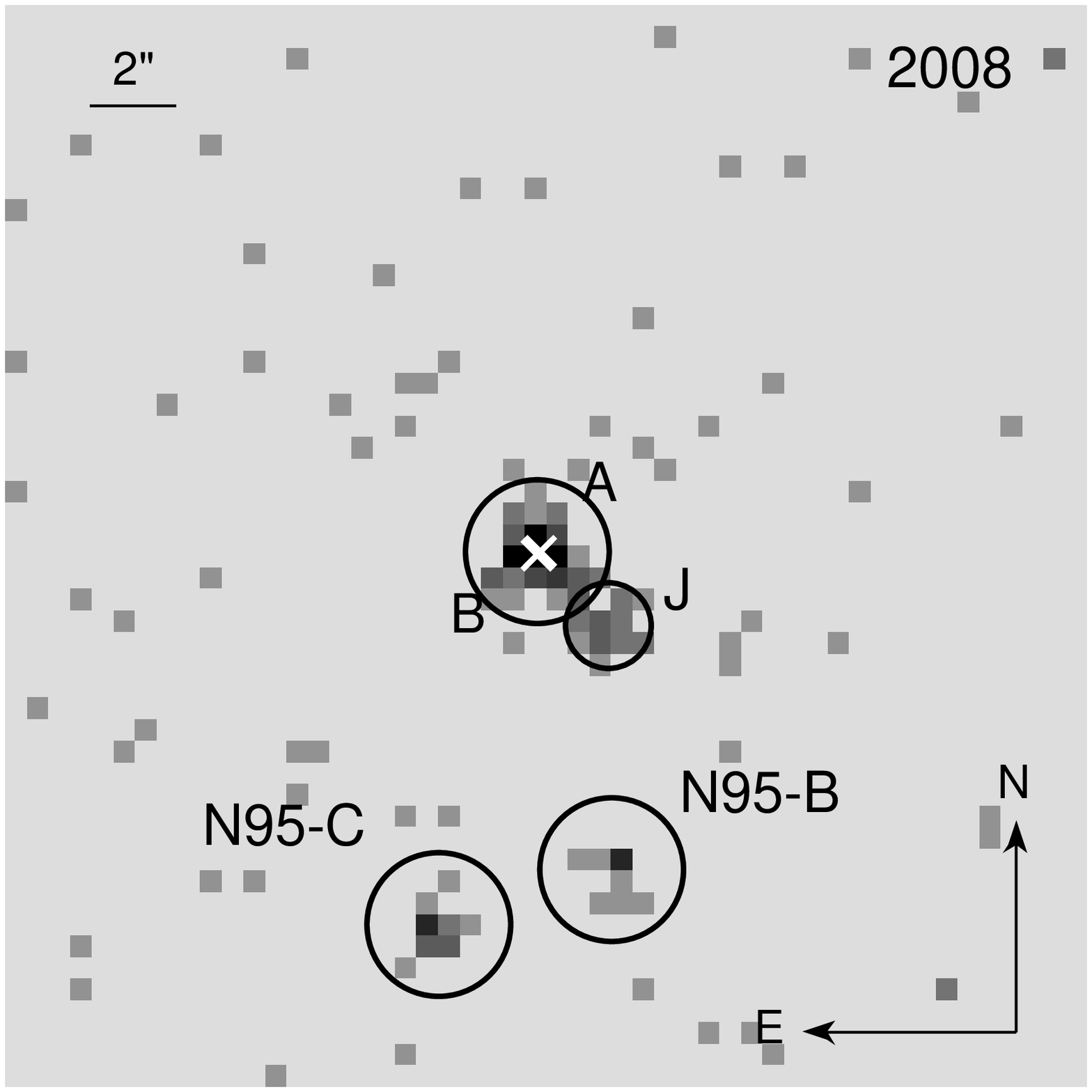}
}
\parbox{2.9cm}{
\vspace*{-1cm}
\hspace*{-1.5cm} \includegraphics[width=3.5cm,clip]{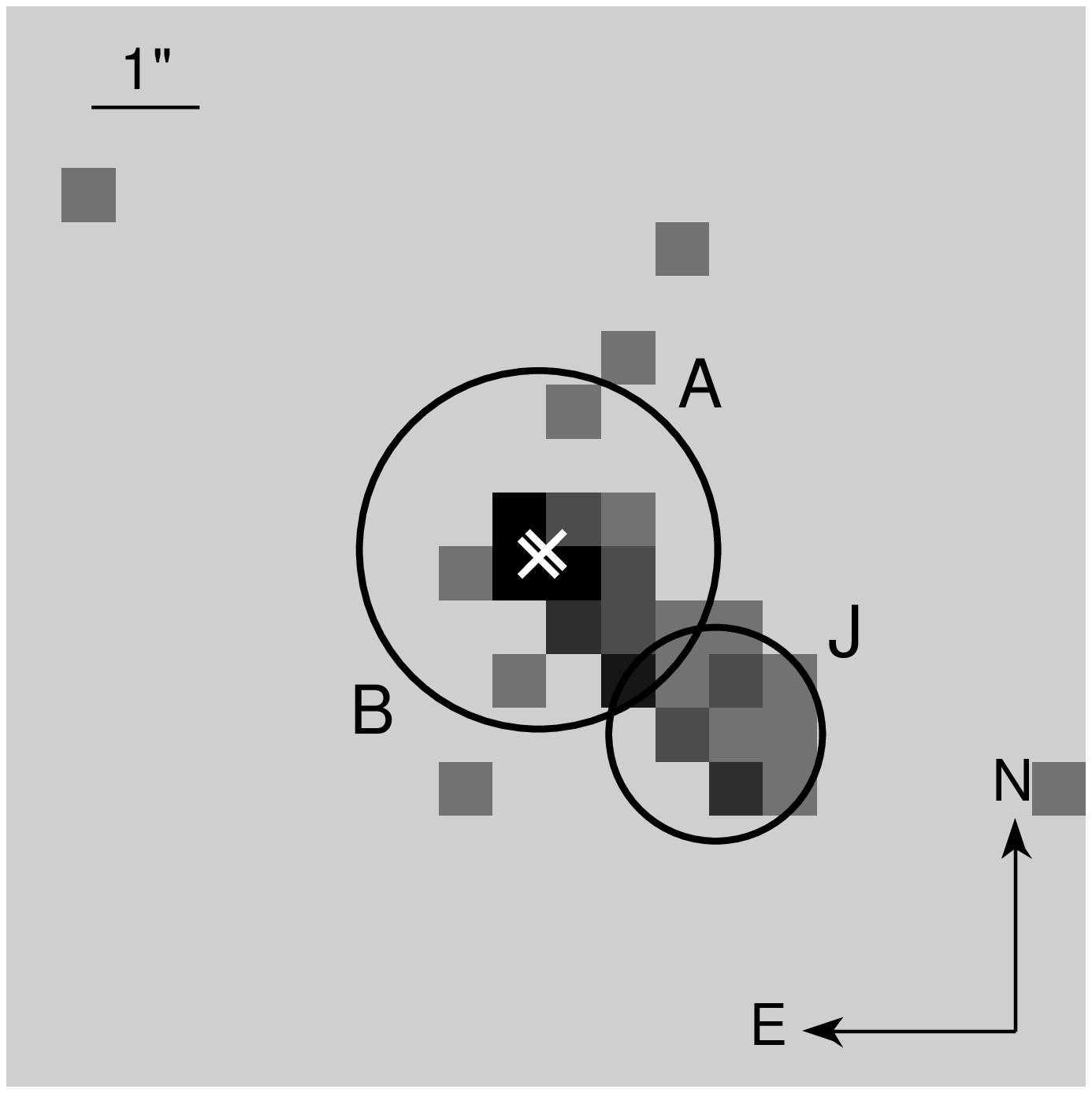}
}
\caption{{\em Chandra}/ACIS images for the $0.2-8$\,keV band in a $25^{\prime\prime} \times 25^{\prime\prime}$ 
region around Z\,CMa ($0.5^{\prime\prime}$ pixels) for the 2003 (top) and the 2008 (bottom) observation; 
zoom on Z\,CMa in the $0.2-1$\,keV band for both epochs. 
All detected X-ray sources are marked with black circles, and the binary components are shown as white x-shaped 
symbols. The position of the X-ray source detected in the 2008 image to the SW of the central source 
labeled `J' is overlaid in the 2003 image as red circle. In 2008, the soft band image shows continuous emission 
along the jet axis rather than distinct X-ray sources.}
\label{fig:acis_images}
\end{center}
\end{figure}

\section{Data analysis and results}\label{sect:data_analysis}

A {\em Chandra} observation of Z\,CMa carried out in Dec 2003 was presented by \cite{Stelzer06.3}
in the framework of a survey for X-ray emission from Herbig stars. During that observation
the source was not in optical outburst. We obtained another 
$40$\,ksec of {\em Chandra} Director's Discretionary Time to search for changes of the X-ray characteristics 
during the recent optical outburst. Table~\ref{tab:obslog} gives the observing log.
\begin{table}
\begin{center}
\caption{Observing log for {\em Chandra} exposures of Z\,CMa.} 
\label{tab:obslog}
\begin{tabular}{rcccc}\hline
\multicolumn{1}{c}{ObsID} & \multicolumn{1}{c}{SeqNo} & \multicolumn{1}{c}{Instrument} & \multicolumn{1}{c}{Date [UT]} & \multicolumn{1}{c}{Expo [ksec]} \\
\hline
3751  & 200221 & ACIS-S2 & 2003-12-07 22:00 & 38 \\
10845 & 200569 & ACIS-S3 & 2008-12-28 22:48 & 40 \\
\hline 
\end{tabular}
\end{center}
\end{table}
For consistency we re-analyze here the 2003 observation analogous to the new data set.

We used the CIAO 
package\footnote{CIAO is made available by the {\em Chandra} X-ray Center and can be downloaded 
from http://cxc.harvard.edu/ciao/download/}, version 4.0, and  
we started with the level\,1 events file provided by the {\em Chandra} X-ray Center. 
In the process of converting the level\,1 events file to a level\,2 events file
for each of the observations we performed the following steps: 
We removed the pixel randomization 
in order to optimize the spatial resolution. 
We filtered the events file for event grades
(retaining the standard grades $0$, $2$, $3$, $4$, and $6$), 
and applied the standard good time interval file. 

For our science goal of detecting the X-ray source(s) associated with Z\,CMa,  
source detection was restricted to a $100 \times 100$ pixels wide image 
(1\,pixel $= 0.5^{\prime\prime}$) and a congruent, monochromatic exposure map for $1.5$\,keV
centered on the optical position of Z\,CMa. 
Source detection was carried out with the {\sc wavdetect} algorithm \citep{Freeman02.1}.
We used wavelet scales between $1$ and $8$ in steps of $\sqrt{2}$. 
We tested a range of detection significance thresholds, and found that $\sigma_{th} = 10^{-5}$ 
avoided spurious detections and at the same time separated close emission components. 

The two X-ray images are shown in Fig.~\ref{fig:acis_images} with optical positions of
the binary and detected X-ray sources overlaid. 
The brightest X-ray source coincides in both observations with the unresolved binary star. 
Two new X-ray sources, not seen in the 2003 data, are detected in the 2008 image. Especially
remarkable is the faint source in the SW elongation of Z\,CMa (named `J' henceforth). The position
angle of this source is $(225 \pm 5)^\circ$, roughly in agreement with the orientation of the blue-shifted part
of the optical jet detected by \cite{Poetzel89.1}. 
In the broad band image ($0.2-8$\,keV) this source is separated by $2.4^{\prime\prime}$ from Z\,CMa.
If only photons below $1$\,keV are considered the 2008 image shows elongated emission along the same axis
suggesting a chain of weak X-ray sources extending to the SW of Z\,CMa. 
There is no significant soft excess in the SW direction during 2003 (see zoom in Fig.~\ref{fig:acis_images}). 
The remaining X-ray sources (labels `N95-B' and `N95-C' in Fig.~\ref{fig:acis_images}) are tentatively identified with 
pre-MS candidates mentioned by \cite{Nakajima95.1}. 
In the following we concentrate on the analysis of the X-ray emission associated with Z\,CMa and its jet.

\begin{table}\begin{center}
\caption{X-ray parameters of the Z\,CMa binary and its jet.}
\label{tab:xrayparams}
\begin{tabular}{lcrrrr}\hline
\multicolumn{1}{c}{Date} & \multicolumn{1}{c}{Opt/IR} & \multicolumn{1}{c}{Offax}       & \multicolumn{1}{c}{Counts}           & \multicolumn{1}{c}{Expo$^*$}  & \multicolumn{1}{c}{$P_{\rm KS}$} \\
                         &                            & \multicolumn{1}{c}{[$^\prime$]} & \multicolumn{1}{c}{in $0.3-8$\,keV}  & \multicolumn{1}{c}{[sec]} &                \\ \hline
Dec 2003 & Z\,CMa & $  1.95 $ & $   47.6 \pm    6.9$ & $  17857.$ &    $  0.54$ \\
Dec 2003 & jet    & $  1.91 $ & $   <9.4$            & $  17857.$ &    $  -   $ \\
\hline
Dec 2008 & Z\,CMa & $  0.30 $ & $  107.2 \pm   10.4$ & $  39594.$ &    $  0.50$ \\
Dec 2008 & jet    & $  0.34 $ & $   20.7 \pm    4.6$ & $  39610.$ &    $  0.32$ \\
\hline
\multicolumn{6}{l}{$^*$ The effective exposure time in the Dec 2003 observation is} \\
\multicolumn{6}{l}{substantially reduced with respect to the duration of the observation} \\
\multicolumn{6}{l}{because the star is near a chip border.} \\
\end{tabular}
\end{center}\end{table}

We calculated the source count rates in the following way: First, the point-spread-function (PSF) 
was computed for each X-ray 
position. A circular source photon extraction region was defined as the area that contains 
$95$\,\% of the PSF. Only for source `J' the extraction region was restricted
to $90$\,\% of the PSF to avoid contamination from the wings of the central source.
The background was
extracted individually from a squared region centered on the source extraction area and 
several times larger than the latter one.
Circular areas centered on the positions of the X-ray sources were excluded from the background
area. The S/N was computed from the counts summed in
the source and background areas, respectively, after applying the appropriate 
area scaling factor to the background counts. In practice, the background is very low (a fraction
of a count in the source extraction area). 
Finally, count rates were obtained using the exposure time at
the source position extracted from the exposure map. 
We have estimated a $95$\,\% confidence upper limit for the count rate 
at the position of `J' in the Dec 2003 observation (marked red in 
Fig.~\ref{fig:acis_images} left) using the algorithm of \cite{Kraft91.1}. 
In Table~\ref{tab:xrayparams} we summarize the relevant X-ray parameters
of the Z\,CMa binary and the source `J' identified with the jet. 

An individual response matrix and auxiliary response were extracted for
the position of each source using standard CIAO tools. 
The spectrum of the brighter source (Z\,CMa) was binned to a minimum of $10$ counts per bin and that of
the fainter one (`J') to $5$ counts per bin.
As mentioned above, the background of ACIS is negligibly low.
We fitted the spectra in the XSPEC\,12.4.0 environment 
with a one- or two-temperature thermal model subject to photo-absorption 
({\sc wabs $\cdot$ apec} and {\sc wabs $\cdot$ [apec + apec]}, respectively). 
For the brighter source (Z\,CMa) the $1$-T fit of the Dec 2008 observation displays substantial residuals slightly
below $1$\,keV, and we resort to the $2$-T model.
The best fit $N_{\rm H}$ is $7^{+5}_{-6} \cdot 10^{21}\,{\rm cm^{-2}}$, 
compatible with the range of values published for the optical extinction 
$A_{\rm V}=2.4 ... 4.6$\,mag \citep{Elia04.1, Acke04.1}. 
Assuming a particle density of $0.3\,{\rm cm^{-3}}$ the galactic absorption at $1$\,kpc amounts to 
$\sim 10^{21}\,{\rm cm^{-2}}$. 
Therefore, Z\,CMa may have some additional absorption related to the star forming environment. 
On the other hand, our measurement does not rule out negligible circumstellar absorption. 
In our best-fit model the soft component ($kT_1=0.4^{+0.7}_{-0.2}$\,keV) dominates over the hard
component ($kT_2 \sim 7.5$\,keV; unconstrained) with an emission measure 
$\log{EM_1}\,{\rm [cm^{-3}]}=53.9^{+1.2}_{-0.6}$ vs. $\log{EM_2}\,{\rm [cm^{-3}]}=53.1^{+0.4}_{-0.3}$.
Due to the large confidence intervals of the spectral
parameters, we determine only a lower limit on the X-ray flux, adopting 
a minimum $N_{\rm H}$ of $\sim 10^{21}\,{\rm cm^{-2}}$
coming up for the interstellar absorption but neglecting any possible contribution from the environment
of the star. We find $f_{\rm x} > 1.6 \cdot 10^{-14}\,{\rm erg/cm^2/s}$ for the $0.5-8.0$\,keV band, corresponding 
to $\log{L_{\rm x}}\,{\rm [erg/s]}> 30.3$ at a distance of $1050$\,pc. 

\begin{figure}
\begin{center}
\includegraphics[width=8cm]{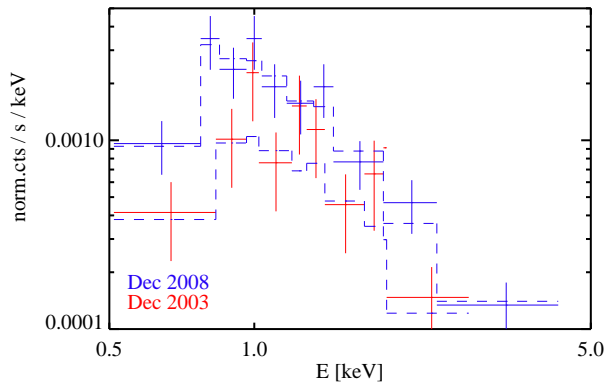}
\caption{Folded {\em Chandra}/ACIS spectra of Z\,CMa. Overplotted on both spectra is the best 
fit model obtained for the 2008 observation (dashed lines).} 
\label{fig:acis_spectra}
\end{center}
\end{figure}

For the Dec 2003 observation due to low photon statistics we can not formally exclude the $1$-T model.
Iso-thermal models are known to be an oversimplification and the number of thermal components needed
to fit stellar X-ray spectra generally increases with photon statistics. We do not present a detailed spectral
analysis of this data set because the quality is poor. However, we can test the compatibility of the 
2003 spectrum with the 2-T model parameters derived from the 2008 observation. 
Indeed, we find from a direct comparison of the 2008 best-fit model to the spectrum observed in 2003, without 
fitting it to the data, a 
$\chi^2_{\rm red} = 1.0$ ($5$ d.o.f.). 
The good agreement is also evident from Fig.~\ref{fig:acis_spectra} where
we plot both observed spectra together with the best-fit model (dashed lines) of the 2008 data. 
Differences in the appearance of the two spectra in Fig.~\ref{fig:acis_spectra} are due to the 
fact that the data and the model are shown folded with the instrument response matrix. 
We recall here that Z\,CMa is located on two different CCD chips and at different off-axis angle in the
2003 and 2008 data sets (see Table~\ref{tab:obslog}), and this implies different spectral response and 
effective area. 

The fainter source (`J') has only $20$ counts in Dec 2008. 
The $1$-T model gives the same absorption as found for Z\,CMa albeit with even larger uncertainties
($N_{\rm H} = 7^{+8}_{-7} \cdot 10^{21}\,{\rm cm^{-2}}$).  
The large error bar of the column density makes it impossible to determine the emission measure and flux
of this source to even an order of magnitude precision.
The temperature of source `J' is also not well constrained 
($0.2^{+0.8}_{-0.1}$\,keV) but probably relatively soft. We find a median photon energy of $0.9$\,keV for
`J' versus $1.2$\,keV for Z\,CMa. 
Analogously to the case of Z\,CMa, we estimate a lower limit for the X-ray flux of `J' 
adopting $N_{\rm H}$ of $\sim 10^{21}\,{\rm cm^{-2}}$ corresponding to the expected 
interstellar absorption. The derived value of $2 \cdot 10^{-15}\,{\rm erg/cm^2/s}$ 
translates to $\log{L_{\rm x}}\,{\rm [erg/s]}> 29.4$ if we assume that the distance of this source is the 
same as for the one associated with Z\,CMa. 
%

Lightcurves were extracted and searched for variability with a maximum likelihood method that divides
the sequence of photons in intervals of constant signal \citep[see ][]{Stelzer07.1} and, independently, 
with the Kolmogorov-Smirnov (KS) test. 
According to this analysis both sources are not variable within the individual exposures. 
The significance level of the KS statistic for each source is given in the last column of Table~\ref{tab:xrayparams}
where the large values of $P_{\rm KS}$ indicate that the data is not significantly different from the assumption
of a constant source.

\section{Discussion}\label{sect:discussion}

\subsection{The central X-ray source}\label{subsect:fuor}

Prior to our observations, only two pre-MS stars have been monitored in X-rays during an 
optical EXor outburst. 
V1647\,Ori has shown a factor $100$ increase of the X-ray luminosity and a steady decay in the subsequent two
years correlated with the near-IR lightcurve. With respect to the pre-outburst state the spectrum was hard
throughout the decaying phase ($3.6$\,keV) and incompatible with X-ray emission from accretion shocks \citep{Kastner06.1}. 
The other EXor object observed in X-rays during an outburst is V1118\,Ori. \cite{Audard05.2} 
showed that its X-ray luminosity varied only by a factor of 
two with respect to the pre-outburst value but the spectrum softened during the EXor event. 
A possible explanation given by
\cite{Audard05.2} was a changing structure of the magnetosphere that moved inward as a result of strong
accretion and disrupted the higher, hotter coronal loops. 
After the outburst V1118\,Ori was seen to have faded by a factor of four in X-ray luminosity but has not 
significantly changed its X-ray temperature \citep{Lorenzetti06.1}.

For Z\,CMa the X-ray brightness and temperature are remarkably constant when 
the 2003 quiesence and the 2008 outburst data are compared. 
Overall, the temperature structure of the X-ray spectrum of Z\,CMa is 
typical of coronal sources observed at similar statistics \citep[e.g.][]{Schmitt90.1}. Clearly, the spectrum is much
softer than
that of V1647\,Ori during its EXor event.  
The low statistics make it impossible to single out eventual contributions from accretion and/or jet shocks.
Similarly, low statistics impeded \cite{Skinner09.1} to constrain the temperature of the soft
component in the X-ray spectrum of the FUor star V1735\,Cyg that they observed outside outburst 
with {\em XMM-Newton}. Their confidence level on the temperature of the soft component of V1735\,Cyg 
is similar to ours ($0.1-1$\,keV) leaving open the question on whether this emission is accretion-related.
We have estimated a lower limit to the X-ray luminosity of Z\,CMa that is compatible with the X-ray 
luminosities reported by \cite{Skinner09.1} for V1735\,Cyg and FU\,Ori. 
As those authors pointed out, these luminosities are at the high end of the $L_{\rm x}$ distribution
of T\,Tauri stars possibly implying that FUor objects have higher than average T\,Tauri mass. 
For Z\,CMa this observation is consistent with the high mass of $3\,M_\odot$ from the literature. 

The binary nature of Z\,CMa leaves room for further speculations on the origin of the X-ray emission: 
(i) The X-rays could be from the less massive star, which is in this case non-variable both in the
optical and X-rays, and the optical variations 
due to changes in scattered light from the primary. 
However, the recent optical photometry has a clear outburst signature and we discard the 
scattered-light scenario. 
(ii) The optical variations may come from the FUor star and 
the X-rays from the embedded primary. This hypothesis is also unlikely as much higher X-ray absorption 
would be expected than is observed. We conclude that probably the same star (the less massive optical component) 
is responsible for both the optical outburst and the X-ray emission but that both phenomena are unrelated. 
Although the existence of a corona can not be taken for granted for an intermediate-mass star, our recent detection of the 
magnetic field of Z\,CMa supports this interpretation (Hubrig et al., 2009, A\&A submitted).

\subsection{The displaced X-ray source}\label{subsect:jet}

A new X-ray source is detected in the 2008 image in the SW elongation of the source associated with the 
Z\,CMa binary. Its position angle agrees with that of the optical jet \citep{Poetzel89.1} and the radio jet 
\citep{Velazquez01.1}
suggesting an interpretation as emission from an internal jet shock. 
According to the strong shock condition \citep{zel66}, our confidence interval of measured 
X-ray temperatures, $0.1...1$\,keV,  
is compatible with shock velocities between $\sim 300...900$\,km/s.  
At a distance of $\sim 2^{\prime\prime}$, corresponding approximately to the location of the new 
X-ray source, in optical forbidden lines the jet was seen to move at $\leq 600$\,km/s \citep{Poetzel89.1}.  
Considering the unknown inclination this is a lower limit for the true velocity of the jet propagation. 
Our estimate for the shock speed is, 
therefore, not incompatible with the expectation that $v_{\rm shock} < v_{\rm jet}$. 
Z\,CMa has a higher mass than most other pre-MS stars with X-ray detected jets \citep[see summary by][]{Bonito07.1}.
The large distance of the displaced X-ray emission from the central star 
is remarkably similar to that of the only other high-mass pre-MS object with 
known X-ray jet, HH\,80/81 \citep{Pravdo04.1}. For the X-ray luminosity we have estimated a conservative
lower limit of $2.6 \cdot 10^{29}$\,erg/s, that could easily be underestimating the intrinsic luminosity by
one order of magnitude or more if the extinction was much higher than the assumed interstellar value. 
Our result is compatible with X-ray luminosities derived for other pre-MS jets \citep[see][]{Bonito07.1}. 
However, all X-ray observations of this object class are hampered by similarly low statistics and, consequently, 
poorly constrained parameters. 

Due to the relatively low X-ray temperature that can be achieved in pre-MS jets 
an image limited to soft photons should better trace the outflow. Indeed, we find continuous soft emission
between Z\,CMa and source `J' in the 2008 image. 
Possibly several knots along the jet axis are producing X-rays at a time. 
A similar finding was reported for DG\,Tau where the X-ray jet was detected both at close ($50$\,AU) and 
larger ($1140$\,AU) separation from the central star \citep{Guedel08.1,Schneider08.1}.  

In 2003 there was no X-ray source at the position of `J'. 
The upper limit to the count rate at this epoch does not exclude X-ray emission at similar levels as 
observed in the 2008 image. However, examination of the photon centroids 
reveals that all counts are in the upper left of the photon extraction area (see Fig.~\ref{fig:acis_images}) 
indicating weak emission 
in the same SW direction as `J' but at a distance of only $\sim 1.2^{\prime\prime}$ from the central source. 
If this emission is interpreted as the head of a moving shock front that has expanded (and brightened)
within the last $5$ years to the position of `J', its projected velocity would be $\sim 900$\,km/s.  
This value is in rough agreement with our measurement from the X-ray temperature. 
However, the analysis of the soft-band image from 2008, where continuously extended soft emission 
from Z\,CMa out to $\sim 2.4^{\prime\prime}$ is detected, suggests that 
`J' represents the temporary radiative loss of 
a localized blob of material impacted by an intermittent jet. The weak emission closer to the star
may represent similar but independent events. 

We add that alternative explanations are perceivable for the X-ray emission displaced from Z\,CMa: 
Scattered stellar light reflected from the outflow cavities was discussed by \cite{Bally03.1}
for the case of HH\,154 but seems unlikely to hold for Z\,CMa because of the large displacement of $> 2000$\,AU
with respect to the central star. 
Again for HH\,154, \cite{Murphy08.1} suggested magnetic reconnection in the space between two interacting 
jets from the two binary components. There is no observed evidence for a second jet in Z\,CMa.
High-quality high-resolution imaging of the Z\,CMa outflow should help to clarify this question. 
Monitoring of the X-ray morphology during the next decade
should also help to constrain the nature of the emission.
For an assumed jet diameter of $100$\,AU we can estimate the electron density from the observed 
X-ray emission measure and derive a radiative cooling time of $\sim 20$\,yrs.


\bibliographystyle{aa} 
\bibliography{zcma}

\end{document}